\documentclass[3p,times]{elsarticle}

\usepackage{ecrc}

\volume{00}

\firstpage{1}

\journalname{Nuclear Physics A}

\runauth{C. Marquet}

\jid{nupha}

\usepackage{amssymb}

\usepackage[figuresright]{rotating}

\usepackage{wrapfig}

\begin{document}

\begin{frontmatter}



\dochead{}

\title{Azimuthal angle and rapidity dependence of di-hadron correlations in QCD}


\author{Cyrille Marquet}

\address{Physics Department, Theory Unit, CERN, 1211 Gen\`eve 23, Switzerland}

\begin{abstract}

I discuss novel QCD phenomena recently observed in p+p, p+A and A+A collisions, that result from the non-linear dynamics of small-x gluons. I focus on di-hadron correlation measurements, as opposed to single-hadron observables often too inclusive to distinguish possible new effects from established mechanisms. Specifically, I discuss angular correlations of forward di-hadrons in d+Au collisions and long-range rapidity correlations in high-multiplicity p+p and Au+Au collisions.


\end{abstract}


\end{frontmatter}


\section{Introduction}

The large amount of experimental data collected at RHIC during the last decade, and at the LHC during the last year, allows to explore novel QCD effects. In particular, RHIC measurements present a number of features which can be consistently interpreted as consequences of a large gluon density in the gold nucleus, and of the corresponding non-linear QCD dynamics. Manifestations of this so-called saturation regime of QCD include the suppression of forward particle production \cite{Arsene:2004ux,Adams:2006uz} and the azimuthal de-correlation of forward hadron pairs \cite{Braidot:2010zh,Meredith:2010zz} in d+Au compared to p+p collisions. The recent start of the LHC significantly enhances the possibilities to explore the saturation regime of QCD already in p+p collisions, as shown by the recent observation of long-range rapidity correlations in high-multiplicity events \cite{Khachatryan:2010gv}.

We argue below that these data are well described by the effective theory called the Color Glass Condensate (CGC), the best approximation of QCD devised so far to quantify parton saturation (see e.g. \cite{Gelis:2010nm}). It properly takes into account both the non-linear evolution of hadronic/nuclear wave functions, and multiple parton interactions in scattering processes. Such effects are relevant when the gluon densities involved are large enough, corresponding to partons with a small-enough energy fraction $x$, and the CGC describes reliably only that small-$x$ part of the hadronic/nuclear wave function. And although the CGC describes non-linear QCD dynamics, it is still a weakly-coupled theory.

In p+A collisions, the saturation regime is better studied with forward particle production, sensitive only to large-x partons on the proton side, but mainly to small-$x$ gluons on the nucleus side. Since the large-$x$ part of a proton wave function is well understood in perturbative QCD, this allows to investigate the small-$x$ part of the nucleus wave function. In A+A and high-multiplicity p+p collisions, the dynamics of small-$x$ gluons can be studied using long-range rapidity correlations, sensitive to the early times after the collision rather than the following complicated space-time evolution of the system. While the existence of the QCD saturation regime is, at a theoretical level, clear, the real challenge from a phenomenological point of view has been to assess to what extent it is relevant at present energies. Such was a difficult task, since different aspect of the parton dynamics may concur in data, and also because the limit of asymptotically high energy in which the CGC formalism is developed is not fully realized in current experiments.

Nevertheless, this has been achieved with the successful simultaneous description of the suppression of particle production \cite{Albacete:2010bs} and azimuthal correlations \cite{Albacete:2010pg} at forward rapidities in d+Au compared to p+p collisions, using the most up-to-date theoretical tools available in the CGC approach. This is explained in Section \ref{Monojets}, in parallel with more standard mechanisms that successfully describe particle production at mid-rapidity. In Section \ref{Ridge}, we discuss long-range rapidity correlations in high-multiplicity p+p and Au+Au collisions, and argue that the angular structure seen in the data and  called the ridge may be interpreted as a consequence of large parton densities \cite{Dusling:2009ni,Dumitru:2010iy}, albeit in these cases several other possible interpretations exist.

\section{Di-hadron correlations, p+p vs d+Au collisions}
\label{Monojets}

In the case of double-inclusive hadron production $pA\!\to\!h_1h_2X$, the partons in the proton (nucleus) wave function that can contribute to the cross section carry a fraction of longitudinal momentum bounded from below by $x_p$ ($x_A$), with
\begin{equation}
x_p=\frac{|p_{1\perp}|\ e^{y_1}+|p_{2\perp}|\ e^{y_2}}{\sqrt{s_{NN}}}\ ,\hspace{0.5cm}
x_A=\frac{|p_{1\perp}|\ e^{-y_1}+|p_{2\perp}|\ e^{-y_2}}{\sqrt{s_{NN}}}\ ,
\label{kinematics}
\end{equation}
where $p_{1\perp},$ $p_{2\perp}$ and $y_1,$ $y_2$ denote the transverse momenta and rapidities of the final-state particles, and $\sqrt{s_{NN}}$ the collision energy per nucleon. One can distinguish three interesting situations, where different QCD dynamics is at play. First, the production of high-$p_T$ di-jets, initiated by large-$x$ partons and kinematically bound to be produced at mid-rapidity. Second, the production of low-$p_T$ di-hadrons at mid-rapidity, sensitive to power corrections and values of $x$ still moderately large. Note that moving one of the hadrons to forward rapidities (central-forward case) increases significantly the value of $x_p$ compared to the central-central case (for which $x_p\simeq x_A \simeq p_T/\sqrt{s_{NN}}$), but decreases $x_A$ only marginally; for this reason we shall refer to these two situations as mid-rapidity ones. Finally, the production of di-hadrons at forward rapidities (forward-forward case) is sensitive to power corrections in the small-$x$ regime, i.e. to saturation. Only the forward-forward case is sensitive to values of $x_A$ as small as in the single-inclusive forward hadron production case: $x_p\!\lesssim\!1$ and $x_A\!\ll\!1$. 

\subsection{High-$p_T$ di-jets and the leading-twist approximation}

\begin{wrapfigure}{r}{0.36\textwidth}
\vspace{-1.3cm}
\begin{center}
\includegraphics[width=5.9cm]{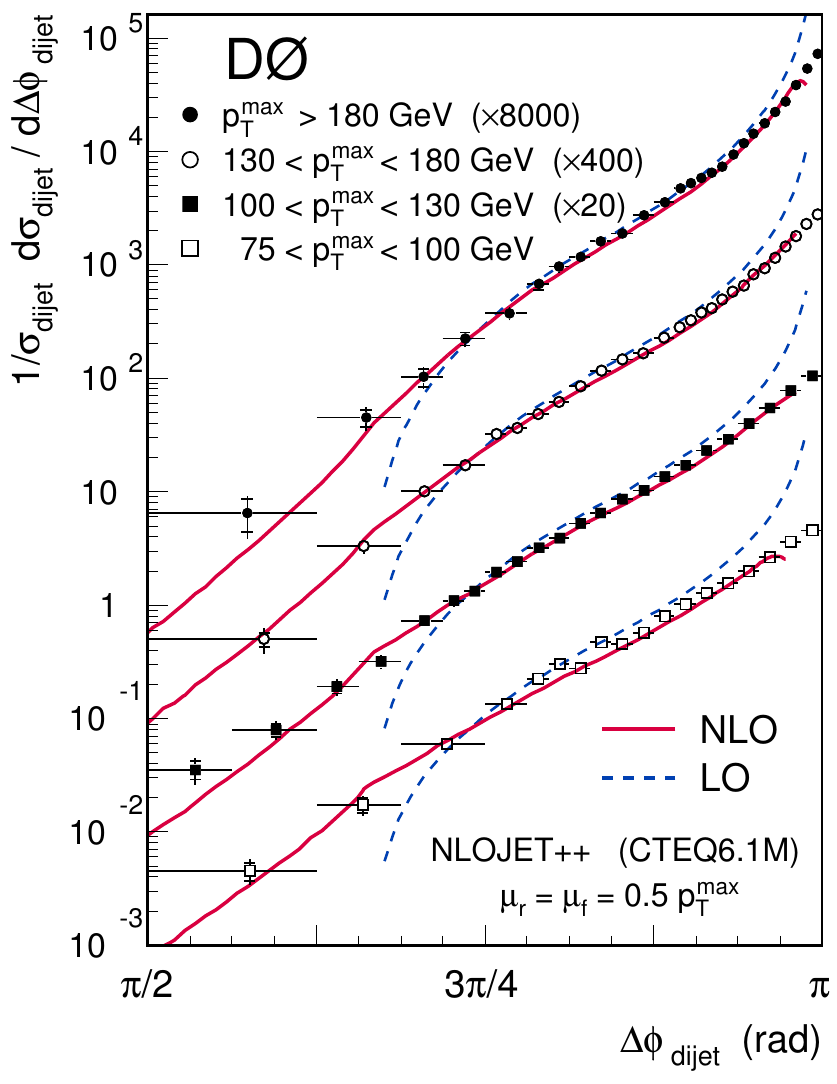}
\end{center}
\caption{High-$p_T$ mid-rapidity di-jet cross-section measured by the D0 collaboration \cite{Abazov:2004hm} as the function of $\Delta\phi$. Although decreasing the $p_T$ range leads to azimuthal decorrelation, the signal remains sharply peaked around $\Delta\phi=\pi$. This is in agreement with leading-twist NLOQCD predictions, which contain up to four partons in the final state.}
\vspace{2cm}
\label{dijets}
\end{wrapfigure}

Fig.~\ref{dijets} shows the cross section for high-$p_T$ di-jets in p+p collisions at the Tevatron, as the function of the azimuthal angle between the jets $\Delta\phi$ \cite{Abazov:2004hm}. The leading-twist approximation of QCD, in which the cross section is computed to leading-power of the hard scale, describes the data very well, and in this regime the di-jet azimuthal correlation is extremely peaked around $\Delta\phi=\pi.$ It is often assumed that this collinear factorization approach also holds in p+A collisions, meaning that highly-virtual partons in nuclei behave independently as they do in protons. Whether or not this is the case could be tested at the LHC, starting with inclusive-jet/hadron measurements \cite{QuirogaArias:2010wh}.

\subsection{Low-$p_T$ di-hadrons and power corrections at large $x$}

\begin{figure}[t]
\begin{center}
\includegraphics[height=4.9cm]{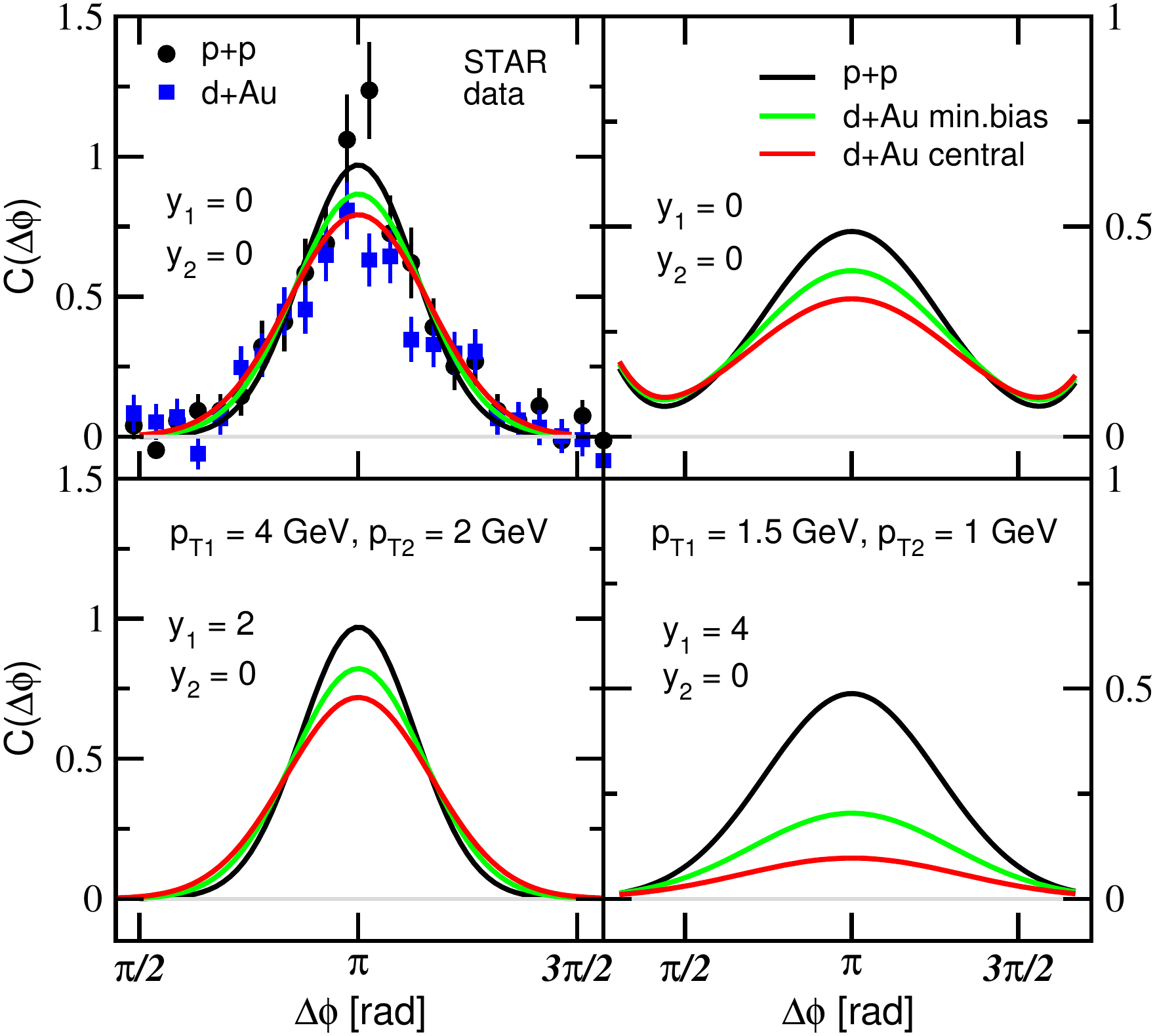}
\hfill
\includegraphics[height=5cm]{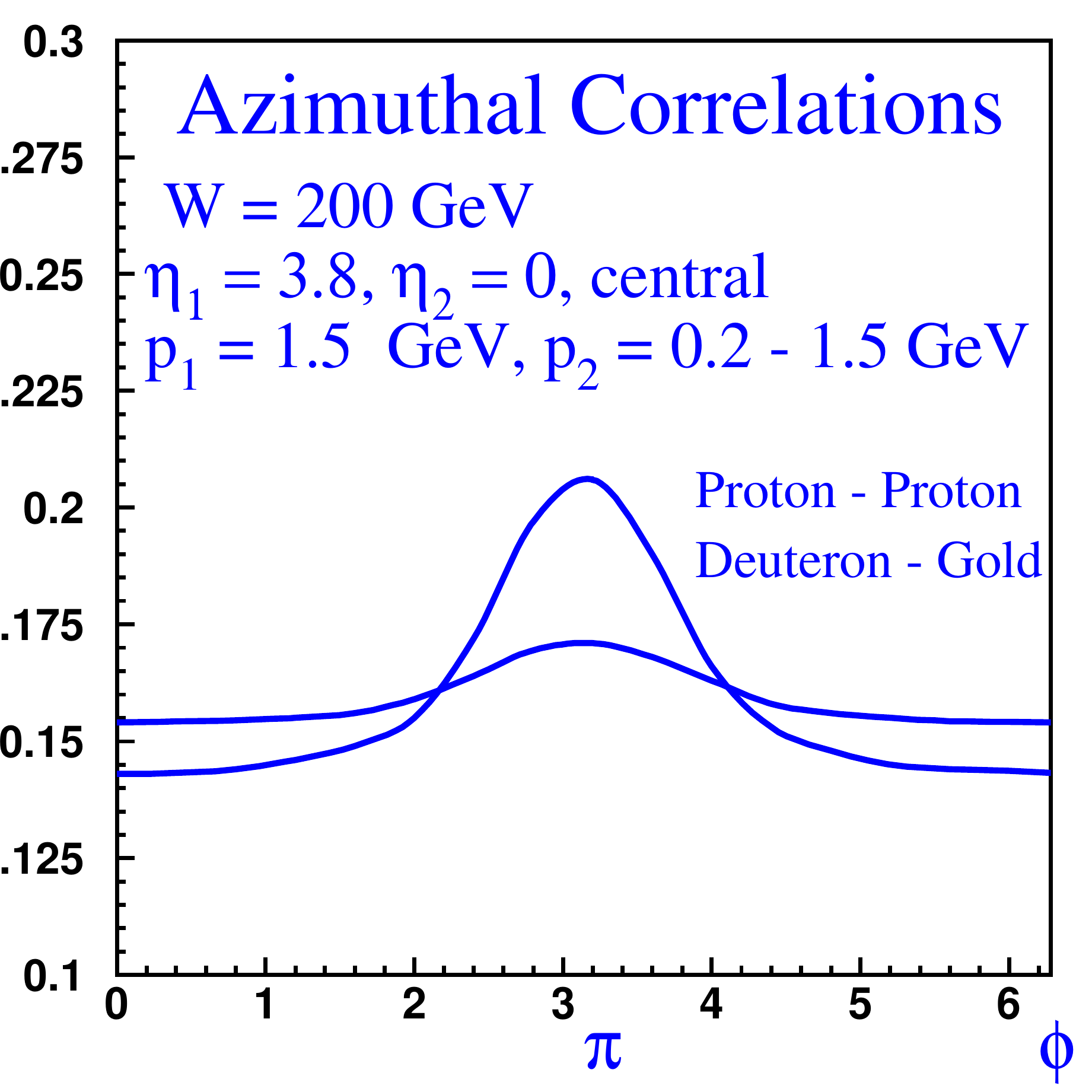}
\hfill
\includegraphics[height=5.1cm]{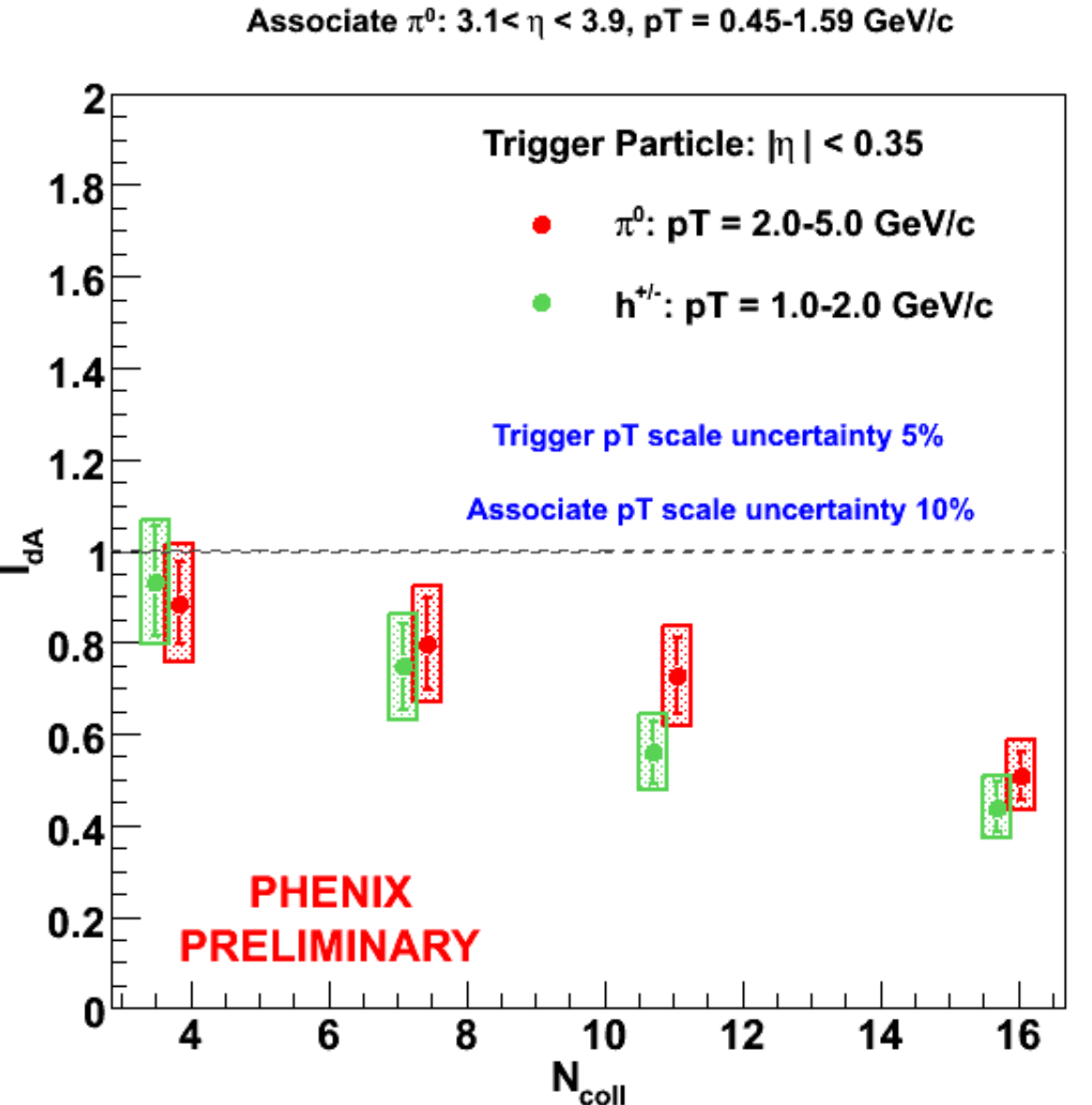}
\end{center}
\caption{The coincidence probability at mid-rapidity as a function of $\Delta\phi$. RHIC data show that in the central-central case the away-side peak is similar in d+Au and p+p collisions. In the central-forward case, the Glauber-eikonal (left plot from \cite{Qiu:2004da}) and CGC (center plot from \cite{Kharzeev:2004bw}) calculations predict that the away-side peak is suppressed in d+Au compared to p+p collisions. This is in agreement with central-forward preliminary $I_{dAu}$ data (right plot from \cite{Meredith:2009fp}), which show that in central d+Au collisions, the integral of the coincidence probability is about half that in p+p collisions, reflecting the depletion of the away-side peak. Collinear factorization cannot reproduce this behavior, while both Glauber-eikonal and CGC calculations predicted it.}
\label{fwd-cent}
\end{figure}

We shall first focus on di-hadron production at mid-rapidity, including both the central-central and central-forward situations, and study the $\Delta\phi$ dependence of the double-inclusive hadron production cross section, where $\Delta\phi$ is the difference between the azimuthal angles of the measured particles $h_1$ and $h_2$. Nuclear effects on di-hadron correlations are typically evaluated in terms of the coincidence probability to, given a trigger particle in a certain momentum range, produce an associated particle in another momentum range.

In a p+p or p+A collision, the coincidence probability is given by $CP(\Delta\phi)=N_{pair}(\Delta\phi)/N_{trig}$, with
\begin{equation}
N_{pair}(\Delta\phi)=\int\limits_{y_i,|p_{i\perp}|}\frac{dN^{pA\to h_1 h_2 X}}{d^3p_1 d^3p_2}\ ,\quad
N_{trig}=\int\limits_{y,\ p_\perp}\frac{dN^{pA\to hX}}{d^3p}\ .
\label{coincproba}
\end{equation}
First measurements were performed at RHIC by the PHENIX and STAR collaborations \cite{Adams:2006uz,Adler:2006hi}. In the central-central case, the coincidence probability features a near-side peak around $\Delta\phi=0,$ when both measured particles belong to the same mini-jet, and an away-side peak around $\Delta\phi=\pi,$ corresponding to hadrons produced back-to-back. In the central-forward case, there is naturally no near-side peak. Either in p+p or d+Au collisions, the sizable width of the away-side peak cannot be described within the leading-twist collinear factorization framework. This indicates that power corrections are important when $|p_\perp|\sim 2$ GeV. At such low transverse momenta, collinear factorization does not provide a global picture of particle production at RHIC, even at mid-rapidity.

There are two main formalisms to take into account power corrections to the leading-twist approximation, the Glauber-eikonal approach, and the CGC framework. The former relies on a resummation of incoherent multiple scatterings, and therefore is not applicable at small $x$, where multiple scatterings are coherent. Performing the complete resummation including energy-momentum conservation is a challenging task, and more often than not a detour to the strict calculation is taken by introducing non-perturbative parameters such as the so-called intrinsic transverse momentum. By contrast, the CGC is valid only in the small-$x$ limit. Coherent multiple scatterings are taken into account along with the equally important non-linear parton evolution, since both parametrically contribute to the same level when a large gluon density is reached, and including one without the other is not consistent. In both frameworks, the multiple scatterings induce a $p_T$-broadening, characterized either by a non-perturbative scale at large $x$, or by the saturation scale at small $x$, as explained in the next section.

Both the Glauber-eikonal \cite{Qiu:2004da} and CGC \cite{Kharzeev:2004bw} approaches qualitatively describe the di-hadron azimuthal correlation data at mid-rapidity, including the broad width of the away-side peak and its depletion in d+Au collisions when going from central-central to central-forward production. This is illustrated in Fig.~\ref{fwd-cent}, left and center plots. Such a depletion does not occur in p+p collisions, it is due to nuclear-enhanced power corrections, and the p+A to p+p ratio of the integrated coincidence probabilities
\begin{equation}
I_{pA}=\frac{\int d\Delta\phi\ CP_{pA}(\Delta\phi)}{\int d\Delta\phi\ CP_{pp}(\Delta\phi)}
\end{equation}
is therefore below unity. In Fig.~\ref{fwd-cent}, right plot, recent PHENIX data on $I_{dAu}$ \cite{Meredith:2009fp} are displayed as a function of centrality.

At the moment, it is not clear whether the physical origin of the $p_T$-broadening and of the azimuthal de-correlation is parton saturation rather than incoherent multiple scatterings. This is probably due to the kinematic window probed in these measurements: for a hadron momentum of $|p_\perp|\sim 2$ GeV, one is sensitive to $x_A\sim 0.01$ to 0.1 at RHIC energies. In this regime different physical mechanisms may concur, and neither underlying assumption of the two different descriptions is completely fulfilled. The approximations made in CGC calculations will apply best in the forward-forward case discussed next.

\subsection{Forward di-hadrons and parton saturation at small $x$}

\begin{figure}[t]
\begin{center}
\includegraphics[height=4.4cm]{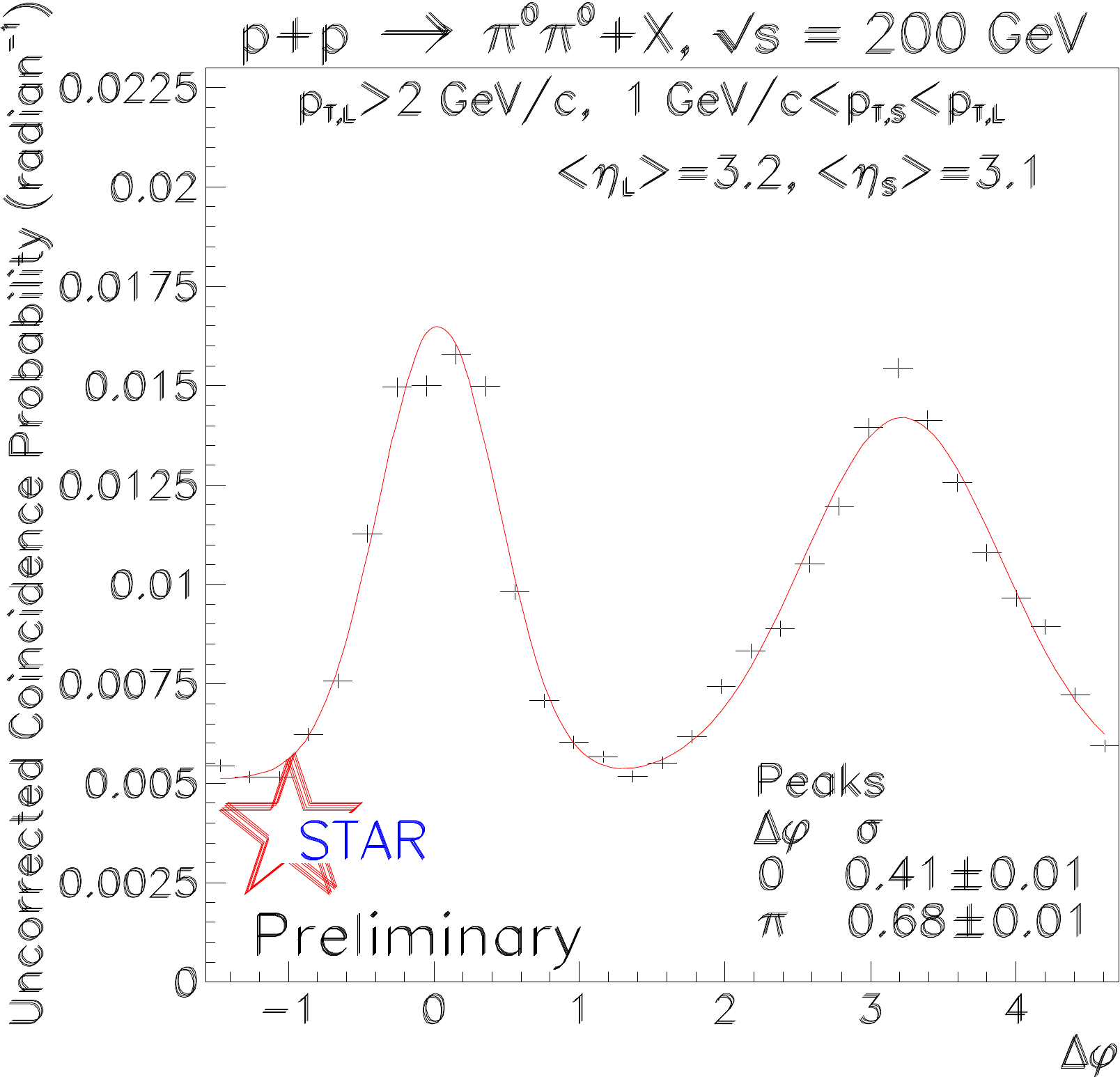}
\hfill
\includegraphics[height=4.4cm]{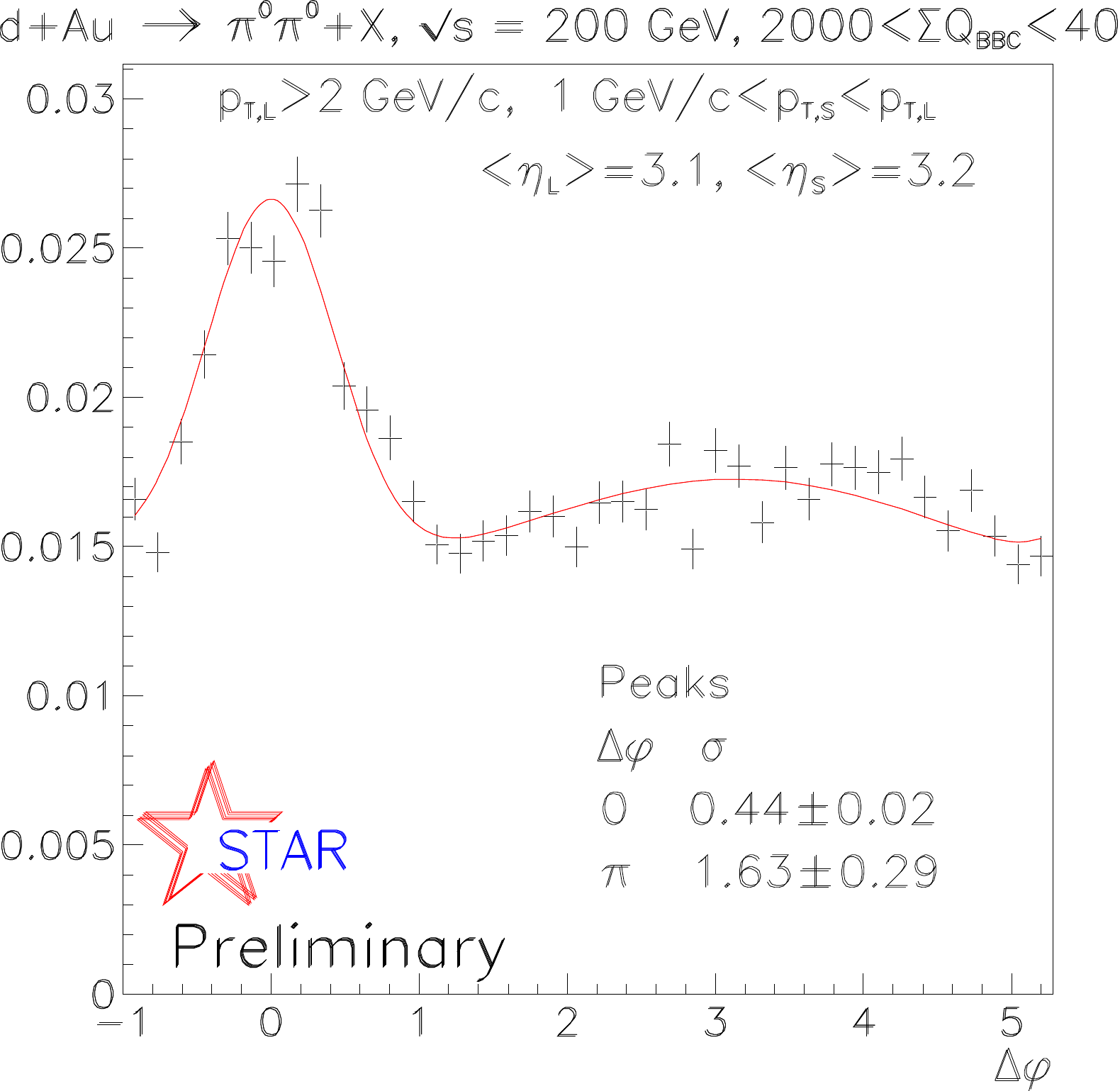}
\hfill
\includegraphics[height=4.4cm]{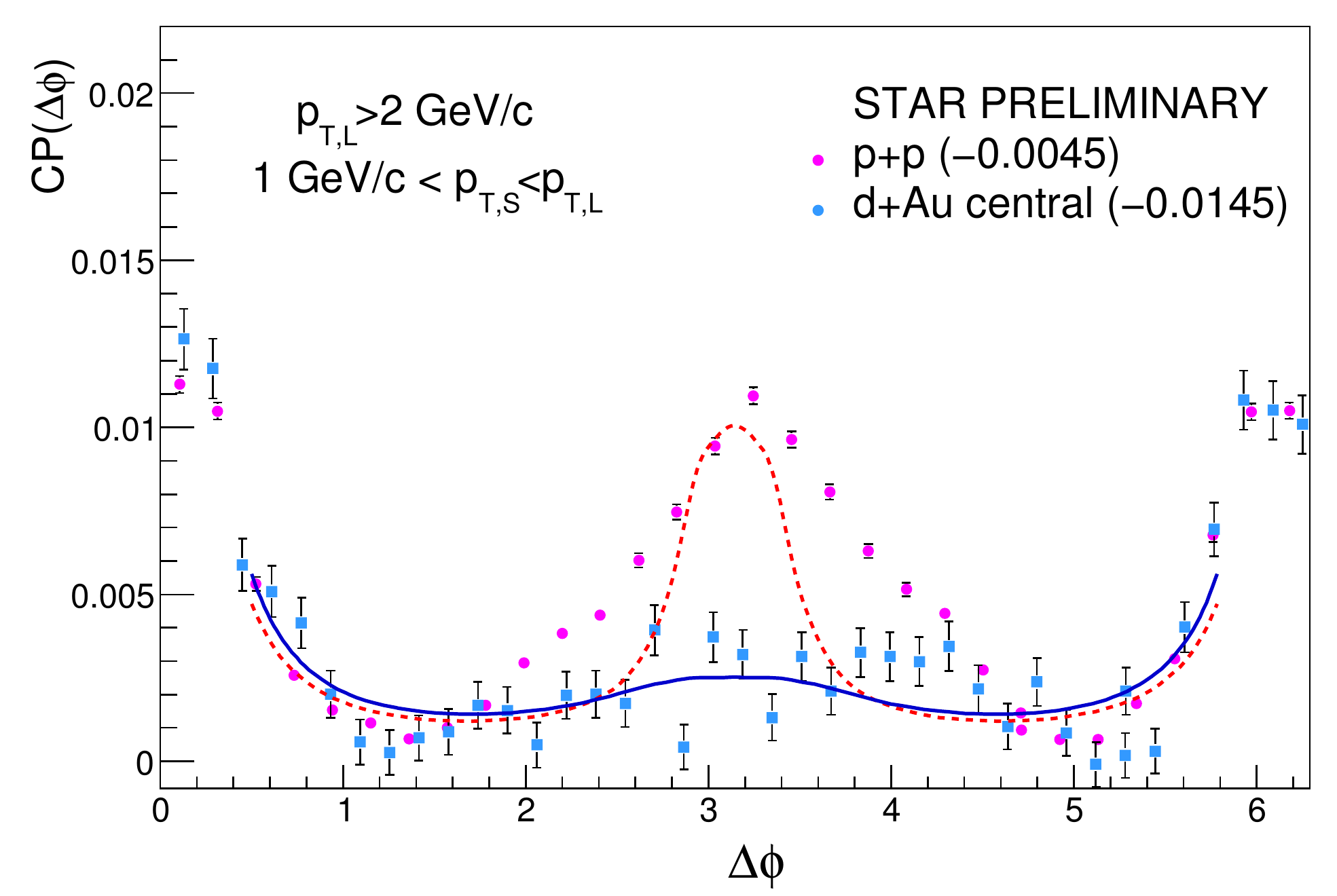}
\end{center}
\caption{The coincidence probability as a function of $\Delta\phi$ for p+p (left) and central d+Au (center) collisions. These preliminary data from \cite{Braidot:2010zh} show a striking nuclear modification of di-hadron azimuthal correlations. The away-side peak, corresponding to hadrons emitted back-to-back, is prominent in p+p collisions but is absent in the central d+Au case. Such production of mono-jets was anticipated in the CGC as a signal of parton saturation, the data are compared with CGC predictions on the right plot; the away-side peak in p+p is qualitatively described by the CGC calculation while its disappearance in central d+Au is quantitatively consistent with the prediction.}
\label{starCP}
\end{figure}

At RHIC mid-rapidity observables are contaminated too much by large-$x_A$ physics to quantitatively explore saturation effects. As explained in the introduction, particle production in the forward-rapidity region offers a cleaner opportunity. This was realized in \cite{Marquet:2007vb}, where the importance of the forward di-hadron measurement was emphasized, and predictions of the azimuthal angle distribution were made. Although these results were only qualitative since for instance parton fragmentation was not included, they have been confirmed by subsequent RHIC measurements. The recent data collected by the STAR collaboration for the coincidence probability obtained with two neutral pions are displayed in Fig.~\ref{starCP}, for both p+p (left plot) and central d+Au (center plot) collisions. The nuclear modification of the di-pion azimuthal correlation is quite impressive, the prominent away-side peak in p+p collisions is absent in central d+Au collisions, in agreement with the behavior predicted in \cite{Marquet:2007vb}.

Let us outline this calculation. As external probes propagate through a target hadron or nucleus, their multiple scatterings off the small-x gluons is coherent because the wavelength of these gluons increases with decreasing x, eventually becoming bigger than the longitudinal extent of the target. The momentum scale below which such coherent multiple scattering are important is the saturation scale $Q_s(x)$, the same scale which characterizes the onset of non-linear effects in the hadronic/nuclear wave function. Both are taken into account in the CGC approach, which resums all power corrections that are dominant in the small-$x$ limit. Non-linear parton evolution generates an intrinsic transverse momentum while some is also gained through multiple scatterings. The magnitude of both is controlled by $Q_s$, which is an increasing function with decreasing $x$.

In the leading-$\log(1/x)$ approximation, the evolution of the saturation scale with $x$, and more generally of the CGC wave function, can obtained from a functional renormalization group equation. In the large-$N_c$ limit, it reduces to the non-linear Balitsky-Kovchegov (BK) equation \cite{Balitsky:1995ub,Kovchegov:1999yj}. The recent determination of running coupling corrections to the original leading-logarithmic equation \cite{Balitsky:2006wa,Kovchegov:2006vj} has proven an essential step in promoting the BK equation to a phenomenological tool. Indeed, the derivation of the rcBK equation, along with numerical simulations \cite{Albacete:2007yr}, made robust quantitative calculations possible. For instance, after the two parameters of the theory were constrained by single-inclusive forward hadron production \cite{Albacete:2010bs} (they are $x_0$, the value of $x_A$ below which one starts to trust, and therefore use, the CGC framework, and the value of the saturation scale at the starting point of the evolution $Q_s(x_0)$), parameter-free predictions for the coincidence probability $CP(\Delta\phi)$ could be made \cite{Albacete:2010pg}.

In Fig.~\ref{starCP}, right plot, these predictions are compared with the STAR data. We see that the disappearance of the away-side peak in central d+Au collisions, compared to p+p collisions, is quantitatively consistent with the CGC calculations. These are only robust in the d+Au case (for which $Q^2_s(x_0\!=\!0.02)=0.6$ GeV$^2$), but the extrapolation to the p+p case (for which $Q^2_s(x_0\!=\!0.01)=0.2$ GeV$^2$ is uncomfortably close to $\Lambda^2_{QCD}$) is displayed in order to show that it is qualitatively consistent with the presence an the away-side peak. Since uncorrelated background has not been extracted from the data, the overall normalization of the data points has been adjusted by subtracting a constant shift, as indicated on the figure.

We recall that di-hadron correlations at mid-rapidity, which are sensitive to larger values of $x_A$, feature an away-side peak in both p+p and central d+Au collisions. The fact that the away-side peak disappears in the central d+Au case, when going from central to forward rapidities, indicates that the effect is correlated with the growth of the nuclear gluon density with decreasing $x_A$. Similarly, we predict that in the forward-forward case, the away-side peak reappears when going from central to peripheral collisions, or for higher transverse momenta of the measured particles. We are not aware of any descriptions of this phenomena that do not invoke saturation effects. We note that another successful description based on the KLN saturation model was also proposed \cite{Tuchin:2009nf}. There, although different assumptions are used, the existence of the saturation scale is the crucial ingredient to successfully reproduce the data.

While more differential measurements of the coincidence probability, as a function of transverse momentum, rapidity or centrality, will provide further quantitative tests of the CGC, this piece of evidence strongly indicates that we have observed manifestations of the saturation regime of QCD at RHIC. However, there are still progresses to be made on the theoretical side. For instance, an approximation was used in \cite{Marquet:2007vb,Albacete:2010pg}, in order to express the double-inclusive hadron production cross section (also derived in \cite{JalilianMarian:2004da,Nikolaev:2005dd,Baier:2005dv} and in \cite{Dominguez:2011wm} with in addition gluon-initiated di-jets) in terms of only the single-gluon distribution function, already well constrained. The validity of this approximation has been critically examined \cite{Dumitru:2010ak}, as it does not allow to correctly implement the non-linear QCD evolution, even in the large $N_c$ limit. Still, in practice, no better solution has been proposed so far. This will require to numerically solve the $x$ evolution of multi-gluon distributions, which implies to introduce additional parameters to characterize the corresponding initial conditions. It was pointed out recently that in the limit of small di-hadron momentum imbalance, only single-gluon distributions are needed \cite{Dominguez:2010xd}, but in general this is not the case.

\section{Long-range rapidity correlations, p+p vs Au+Au collisions}
\label{Ridge}

Let us now discuss high-multiplicity p+p and heavy-ion collisions. When considering low-$p_T$ particle production, only the small-$x$ part of the wave functions of the colliding objects is relevant, and therefore such collisions can be described by the scattering of two CGCs. The system created immediately after such a collision is called the Glasma, it provides a weak-coupling description of the early stages after a high-multiplicity p+p or A+A collision, before evolving into a thermalized quark-gluon plasma it the system size is big enough, as is the case for not-too-peripheral heavy-ion collisions. 

If we denote $\rho_1$ and $\rho_2$ the strong color charges of the incoming hadrons/nuclei, the color field describing the dynamics of the small-x gluons is solution of the Yang-Mills equations
\begin{equation}
[D_\mu,F^{\mu\nu}]=\delta^{+\nu}\rho_1+\delta^{-\nu}\rho_2\ .
\end{equation}
The Glasma field, after the collision, is non-trivial \cite{Lappi:2006fp}: it has a strong component ($A^\mu\sim1/g_s$), a component which is particle like ($A^\mu\sim1$), and components of any strength in between. To understand how this pre-equilibrium system thermalizes, one needs to understand how the Glasma field decays into particles. Right after the collision, the strong field component contains all modes. Then, as the field decays, modes with $p_T>1/\tau$ are not part of the strong component anymore, and for those a particle description becomes more appropriate. After a time of order $1/Q_s$, this picture breaks down, and it has been a formidable challenge to determine weather a fast thermalization can be achieved within this framework.

\begin{figure}[t]
\begin{center}
\includegraphics[width=7.5cm]{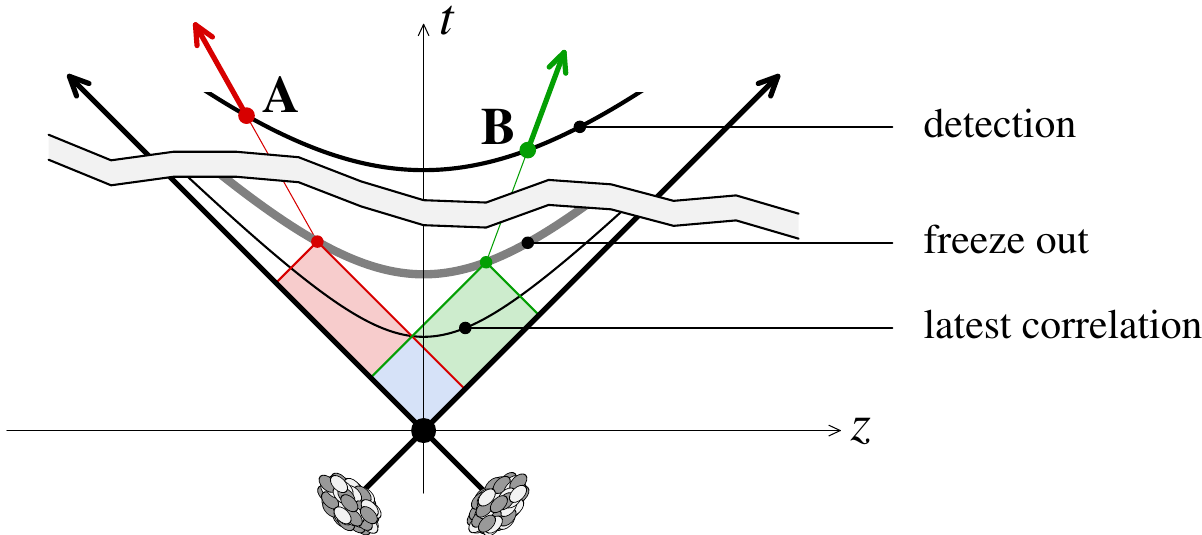}
\hspace{0.5cm}
\includegraphics[width=6.5cm]{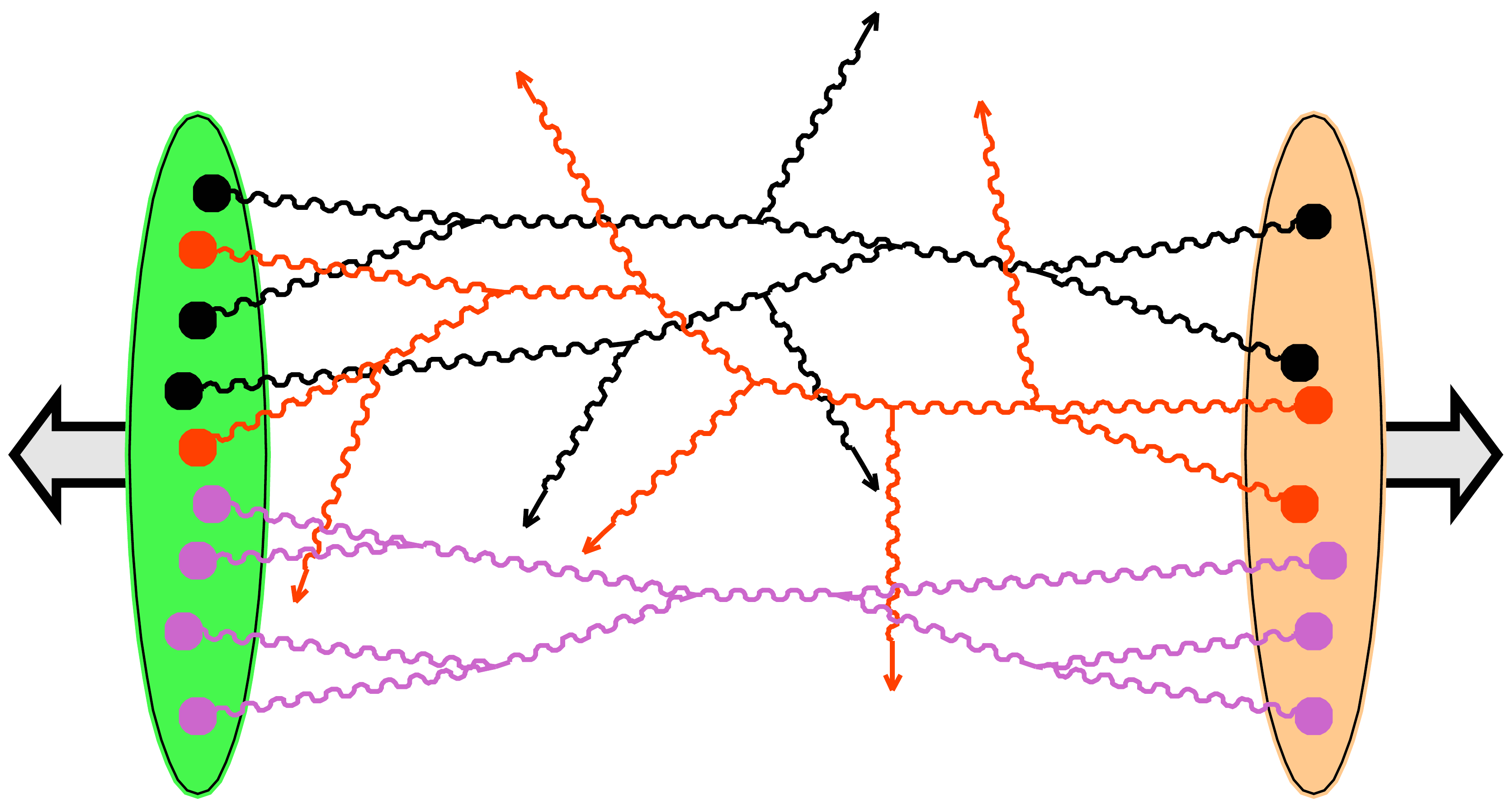}
\end{center}
\caption{Left: the space-time location of events that may correlate two particles is the intersection of their past light-cones. Correlations between particles widely separated in rapidity are due to early-time dynamics. Right: typical leading-order diagram for particle production in the Glasma, multiple partonic interactions are crucial when low values of $x$ are being probed in the nuclear wave functions.}
\label{glasma}
\end{figure}

A problem which can be addressed more easily is particle production. As discussed below, a factorization framework has been established to perform such calculations \cite{Gelis:2008rw}. Before we outline it, let us point out that two-particle correlations are especially interesting, because they can probe the dynamics of the early times after the collision, if the rapidity difference between the two final-state particles is large enough \cite{Dumitru:2008wn}. Indeed, causality imposes that particles separated by a rapidity $\Delta\eta$ can be correlated only by events which happened at
\begin{equation}
\tau<\tau_{f.o.}\ e^{-\Delta\eta/2}\ ,
\end{equation}
where the freeze-out time $\tau_{f.o.}$ denotes the time of last interaction. This is explained in Fig.~\ref{glasma}, left diagram, which shows the space-time picture of the collision: in the forward light-cone, lines of constant proper time $\tau=\sqrt{x^+x^-}$ are hyperbolae and lines of constant rapidity $\eta=\frac12\log(x^+/x^-)$ are straight lines from the origin. Therefore, while heavy-ion collisions are in general not great probes of parton saturation due to the complicated space-time evolution of the system, long-range di-hadron correlations are particularly well-suited observables for that purpose.

\subsection{Particle production in the Glasma}

To compute particle production in the collision of two CGCs, the first task is to express the cross sections in terms of the Glasma field $A_\mu[\rho_1,\rho_2]$, taking into account multiple partonic interactions, as pictured in Fig.~\ref{glasma}, right diagram. Then, one needs to average the cross sections over all possible field configurations, distributed according to functional probabilities that we denote $W_{x_A}[\rho_1]$ and $W_{x_B}[\rho_2]$:
\begin{equation}
\langle O\rangle = \int D\rho_1 D\rho_2 W_{x_A}[\rho_1] W_{x_B}[\rho_2]\ O[A_\mu]\ .
\end{equation}
This factorization has been recently proven \cite{Gelis:2008rw} to leading order in $\alpha_s$, and to all orders in $g_s\rho_1$ and $g_s\rho_2$. The small-x evolution is contained in the CGC "wave functions" $W_x$, while the momenta of the final-state particles enter in the observables $O$. For instance, in the case of single-inclusive gluon production, the cross section reads
\begin{equation}
\frac{dN}{d^2p_\perp dy}[A_\mu]=\frac{1}{16\pi^3}\int d^4x d^4y\ e^{ip\cdot(x-y)}\ \Box_x\Box_y
\sum_\lambda \epsilon_\lambda^\mu(p) \epsilon_\lambda^\nu(p)\ A_\mu(x)A_\nu(y)\ .
\end{equation}

An aspect of hadronic collisions which happens to be much simpler when large parton densities are involved, compared to small ones, is multi-gluon production. Indeed, in a collision of dilute objects of small partonic densities, the produced gluons predominantly all come from the same ladder, because this is what involves the parton distributions a minimum number of times (once for each colliding hadron). By contrast, in a collision of dense objects, the dominant contribution comes about when the produced gluons all originate from different ladders, so as to involve both gluon distributions a maximum number of times (a pair for each produced gluon). The multi-gluon production cross section then reads:
\begin{equation}
\frac{dN}{d^3p_1\dots d^3p_n}[A_\mu]=\frac{dN}{d^3p_1}[A_\mu]\times\dots\times\frac{dN}{d^3p_n}[A_\mu]\ .
\end{equation}
This power counting argument that qualitatively distinguishes multi-particle production in high-multiplicity p+p and A+A collisions compared to minimum-bias p+p collisions, also applies when $\Delta\eta$ is large.

In the leading-$\log(1/x)$ approximation, the evolution of $W_x[\rho]$ with decreasing $x$ is given by a functional renormalization group equation. As both the CGC wave function $W_x$ and the Glasma field $A_\mu$ first have to be evaluated numerically, performing the $\rho$ integrations while fully implementing the non-linear QCD evolution has proven to be an extremely challenging task, and has not been achieved yet. In phenomenological studies of two-particle correlations, it is instead assumed that $W_x[\rho]$ is a Gaussian distribution, whose variance is evolving with $x$ such that the evolution of the single-gluon distribution is correctly reproduced (but that of multi-gluon distributions is not). In this case, the double-inclusive hadron production cross section can be expressed in terms of only single-gluon distributions \cite{Dusling:2009ni}, allowing a practical phenomenological analysis. However, similarly to what was mentioned at the end of Section \ref{Monojets}, this approximation does not correctly implement the non-linear QCD evolution, as it misses a large number of leading-$N_c$ contributions \cite{Dumitru:2010mv}. It was also argued recently that, as a consequence of using Gaussian distributions, two-particle correlations are underestimated \cite{Kovner:2010xk}.

\subsection{The ridge in p+p vs Au+Au collisions}

\begin{figure}[t]
\begin{center}
\includegraphics[width=5.5cm]{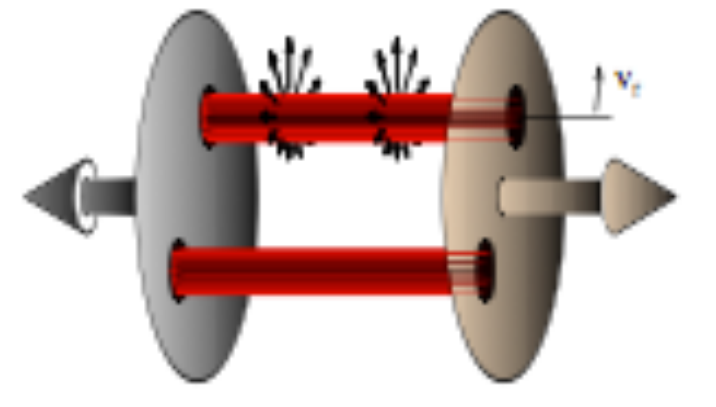}
\hfill
\includegraphics[width=5cm]{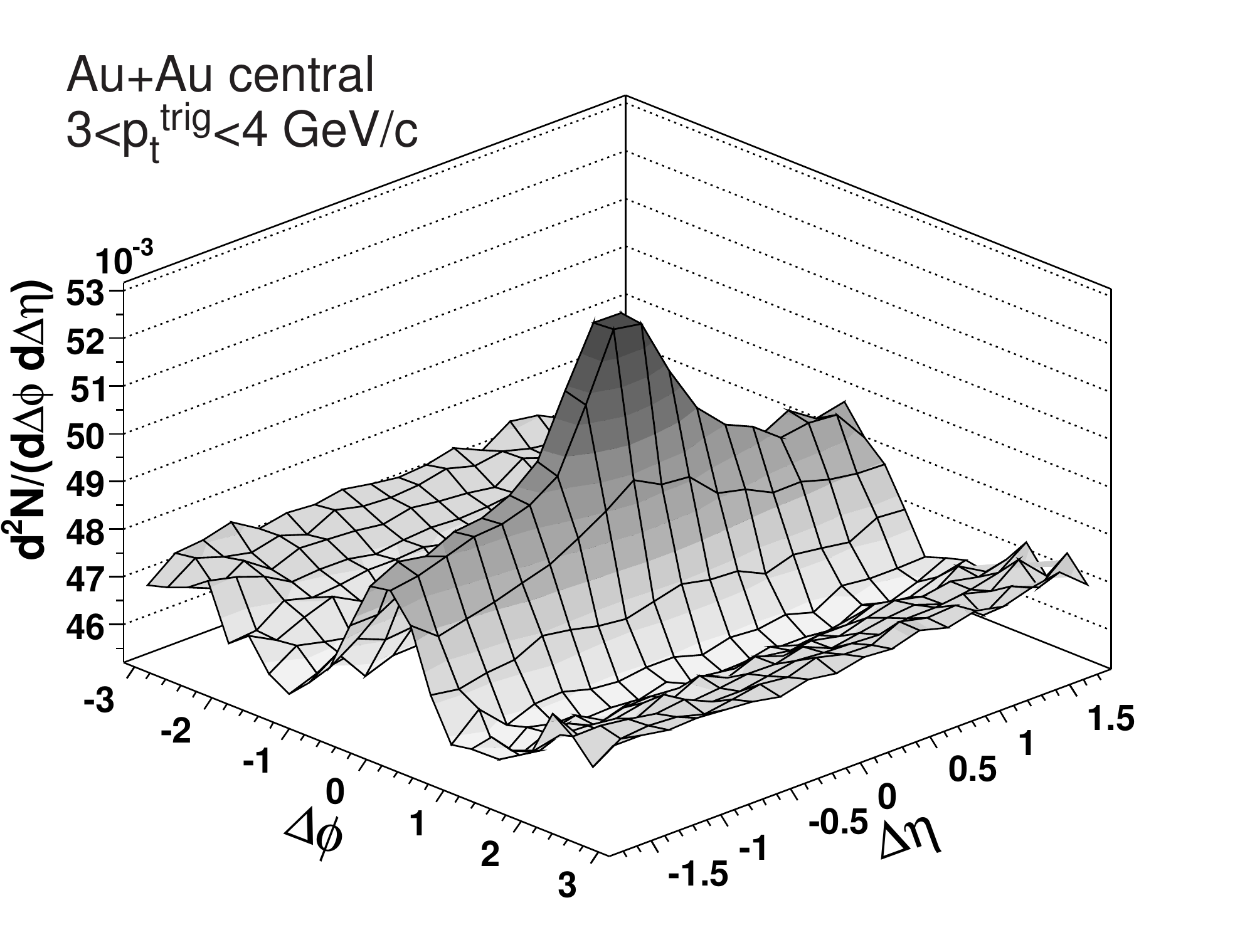}
\hfill
\includegraphics[width=5.5cm]{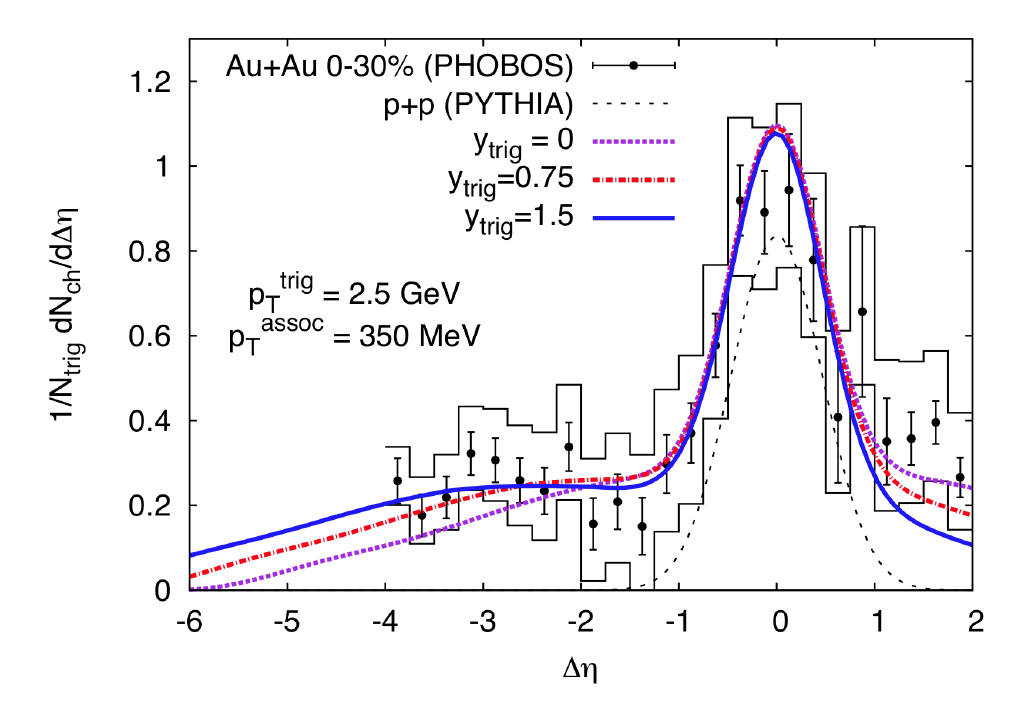}
\end{center}
\caption{Left: figure showing that particles emitted from the same flux tube are correlated in rapidity in a nearly boost-invariant way, and explaining how radial flow turn this a priori $\Delta\phi$-independent correlation into a ridge structure. Center: STAR measurement of di-hadron correlations in central Au+Au collisions \cite{:2009qa}, featuring long-range rapidity correlations on both sides of the jet peak, in qualitative agreement with the CGC expectations. Right: comparison between
PHOBOS data \cite{Alver:2009id} and GCG calculations \cite{Dusling:2009ni}, of the magnitude of the ridge at $\Delta\phi=0$ as a function of $\Delta\eta$.}
\label{ridgeAA}
\end{figure}

Two of the main features of the Glasma field are that, at leading-order it is $\eta$-independent, and that its correlation length in the transverse plane $1/Q_s$ is much smaller than hadronic/nuclear radii. Consequently, the color field $A_\mu$ is made of boost-invariant longitudinal flux tubes of transverse size $1/Q_s$. Because of this flux-tube structure, the Glasma is a natural candidate to explain the ridge-shaped two-particle correlations observed in Au+Au collisions at RHIC \cite{:2008cqb,Alver:2009id,:2009qa}. This is illustrated in Fig.~\ref{ridgeAA}, left drawing: the particles can come from different tubes in which case they are uncorrelated, or come from the same tube which creates an $\eta$-independent correlation.

Due to the strong radial flow induced at a later stage by the collective motion of the system, this correlation is turned into a ridge collimated around $\Delta\phi=0$, such as the one displayed in Fig.~\ref{ridgeAA}, center plot, obtained from a di-hadron correlation measurement in central Au+Au collisions. In agreement with our interpretation, this ridge-shaped structure is only observed with particles of low-enough transverse momenta, that participate in the collection motion. It disappears in peripheral collisions as well, since these do not create a quark-gluon plasma, and therefore any radial flow. In this case, only a strong-enough intrinsic $\Delta\phi$ dependence of the two-particle correlations could reveal the long-range rapidity correlations.

The right plot of Fig.~\ref{ridgeAA} shows the magnitude of the ridge as a function of $\Delta\eta$, for $\Delta\phi=0$. The data are successfully compared with a calculation that combines an estimate of the jet peak around $\Delta\eta=0$, with the nearly $\eta$-independent correlation computed from the double-inclusive particle production cross section in the CGC \cite{Dusling:2009ni}. The figure also shows that the ridge is several units long in rapidity, confirming that these correlations originate at early times.  

\begin{figure}[t]
\begin{center}
\includegraphics[width=4.5cm]{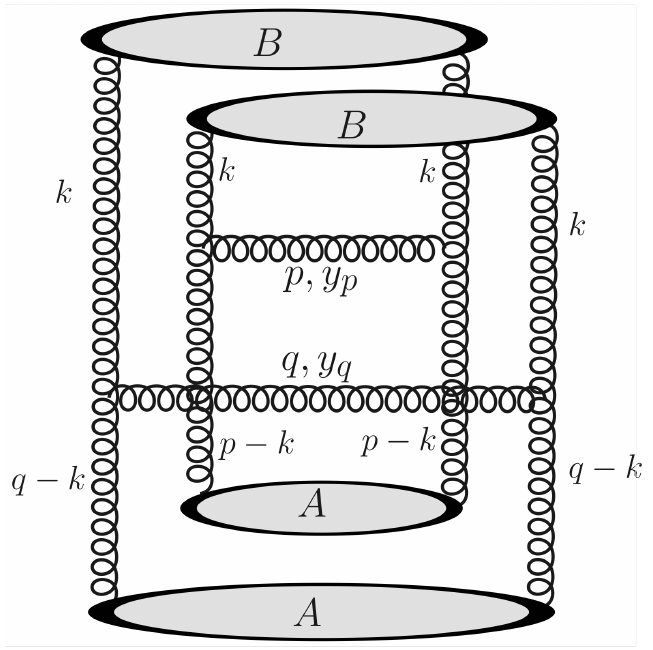}
\hfill
\includegraphics[width=5.5cm]{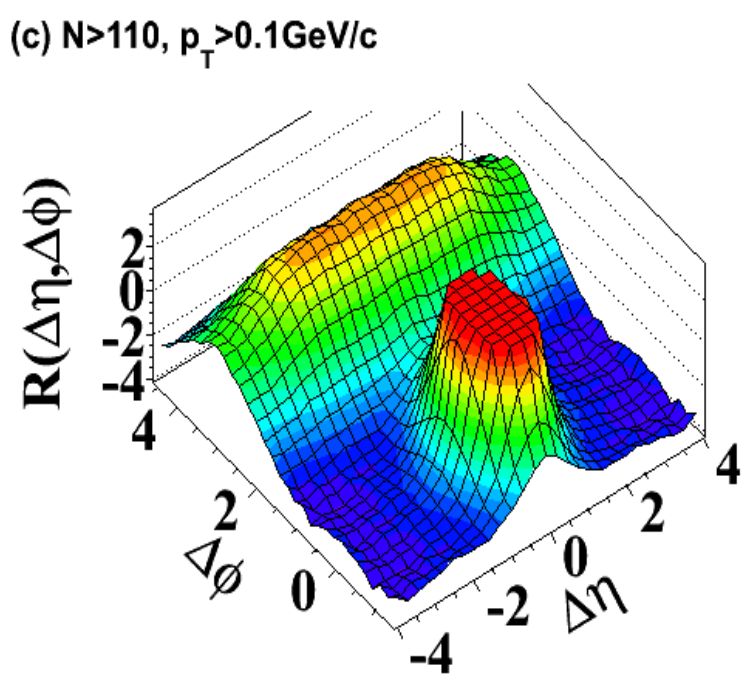}
\hfill
\includegraphics[width=6cm]{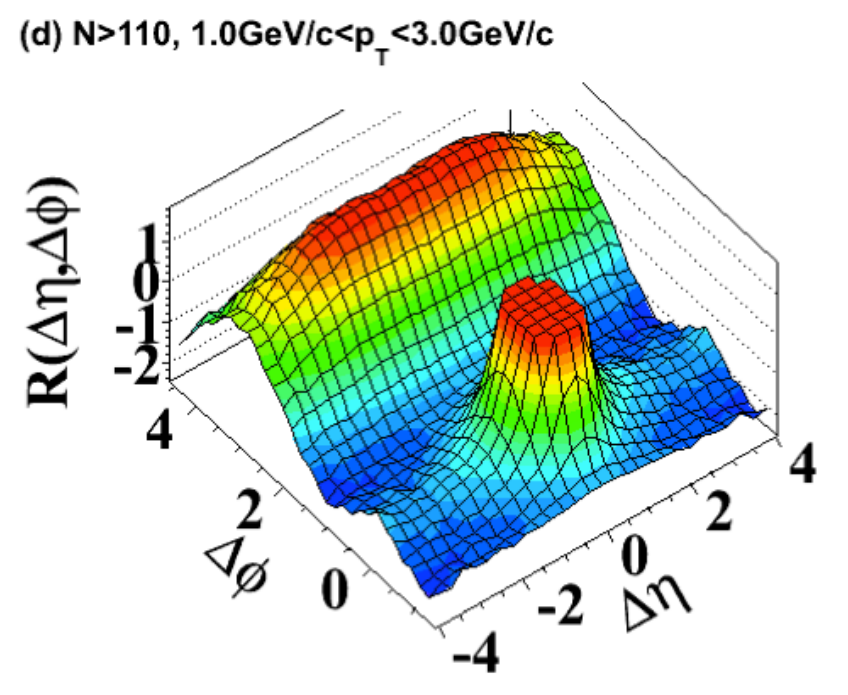}
\end{center}
\caption{Left: a typical diagram that generates an angular collimation around $\Delta\phi=0$ nearly independent of $\Delta\eta$, the rapidity separation between the two hadrons. However, it is non-zero only for hadron transverse momenta comparable to $Q_s$, and it is dominant only at large $\Delta\eta$ away from the jet peak. Center and right: CMS measurement of di-hadron correlations in high-multiplicity p+p collisions \cite{Khachatryan:2010gv}, as a function of $\Delta\phi$ and $\Delta\eta$. The data is consistent with the CGC picture, as the so-called near-side ridge is only observed when $1<p_T<3$ GeV.}
\label{ridgepp}
\end{figure}

Using high-multiplicity events in p+p collisions at the LHC, the CMS collaboration \cite{Khachatryan:2010gv} recently answered the question whether, in a system without much collective flow, one could detect the rather-small intrinsic $\Delta\phi$ dependence of di-hadron correlations in QCD at high energies. It happens to be in qualitative agreement with CGC expectations \cite{Dumitru:2010iy}. For instance, the left diagram of Fig.~\ref{ridgepp}, which involves two (A,B) pairs of gluon distributions as previously explained, is responsible for a $\Delta\phi$ dependence of two-particle correlations, when the transverse momentum of both final-state hadrons is of order $Q_s$. This QCD correlation is irrelevant in heavy-ion collisions, where it is overwhelmed by radial flow, however it dominates when the system is too small to develop any collective motion. This seems to be the case in high-multiplicity p+p collisions at the LHC, where no ridge is seen for low-$p_T$ particles (Fig.~\ref{ridgepp}, center plot). On the contrary, the ridge is indeed present when $p_T\sim Q_s$ (Fig.~\ref{ridgepp}, right plot).

\section{Conclusions}

To conclude, let us emphasize that saturation-based approaches were the only ones to correctly predict the suppression of forward particle production \cite{Kharzeev:2003wz,Albacete:2003iq} and the azimuthal de-correlation of forward hadron pairs \cite{Marquet:2007vb} in p+A vs p+p collisions at RHIC, as well as the ridge phenomenon in high-multiplicity p+p collisions at the LHC \cite{adrian}. While alternative, subsequent-to-data and distinctive explanations of these different phenomena have been and may still be proposed, we are not aware of any formalism that can describe all of them in a global way, other than the CGC, which, let us not forget, is also known to describe well all small-x sensitive observables in e+p collisions at HERA.

\end{document}